\documentstyle[sprocl]{article}

\def\be{\begin{equation}}
\def\ee{\end{equation}}
\def\bea{\begin{eqnarray}}
\def\eea{\end{eqnarray}}

\begin{document}

\hfill{UM-P-96/85}

\hfill{RCHEP-96/10} 

\title{NEUTRINO-ANTINEUTRINO ASYMMETRY IN THE EARLY UNIVERSE}

\author{ R.R. VOLKAS }

\address{Research Centre for High Energy Physics, School of Physics,\\
University of Melbourne, Parkville 3052, Australia}


\maketitle\abstracts{In this talk I discuss the use of a relic
neutrino-antineutrino asymmetry in suppressing active-sterile neutrino
oscillations in the early universe. This phenomenon can serve to
greatly loosen the well known Big Bang Nucleosynthesis constraints on
active-sterile neutrino mixing, and is thus important in some proposed
solutions to the solar and atmospheric neutrino anomalies which employ
large-angle active-sterile mixing. I first discuss the necessary
conditions for the survival of a pre-existing neutrino asymmetry. I
then consider the much more interesting phenomenon whereby such an
asymmetry can actually be created by active-sterile neutrino
oscillations.}

\section{Introduction}

There are three experimental indications for neutrino oscillations:
the solar neutrino problem, the atmospheric neutrino problem and the
LSND experiment. Each is fraught with concerns about the veracity of the
claimed effect. The solar problem is perhaps the most solidly
established, given that the two gallium based experiments GALLEX and
SAGE are technologically impressive as well as because they measure
neutrinos from the $pp$ chain in which theoretical solar model
uncertainties are quite small ($10\%$). The status of the atmospheric
problem will presumably soon be settled by Superkamiokande, while we
will have to wait $2-3$ years to see the LSND claim either confirmed
or disconfirmed by KARMEN\cite{expts}.

If all three effects are real, then at least one species of sterile
neutrino must exist (if the explanation is to lie solely within the
ambit of neutrino oscillations). If one sterile species exists, then
there are strong reasons for thinking that two more such species
should also exist.\cite{epm} 
One is thus led to consider scenarios involving
three active and three sterile neutrino flavours. The solar problem
would then be solved by the mixing of $\nu_e$ with one of the sterile 
species $\nu'_e$, either through the MSW effect or through vacuum
oscillations. Another of the sterile species $\nu'_{\mu}$ can then
usefully be employed to solve the atmospheric problem through
large-angle $\nu_{\mu}-\nu'_{\mu}$ mixing with $\delta m^2 \simeq 10^{-2}$
eV$^2$.\cite{f}$^,$\cite{epm}

This scenario can pose a severe cosmological problem, however. The 
parameters needed to solve the atmospheric problem via active-sterile
oscillations will lead to $\nu'_{\mu}$ being brought into thermal
equilibrium prior to the Big Bang Nucleosynthesis (BBN) epoch, {\it
provided that there is no significant neutrino-antineutrino asymmetry
in the background plasma.} This means that the number of neutrino
species contributing to the expansion of the universe during BBN will
be at least equal to four. This is marginally allowed according to
current BBN analyses.\cite{olive} 
However, some models of ``sterile'' neutrinos
actually have significant self-interactions between them, even though
their interactions with ordinary matter are extremely small. If these
self-interactions are sufficiently strong, then the excitation of one
sterile species will lead to the thermal equilibration of all three
sterile species, thus presenting a clear BBN problem. In addition, the
simple maximal mixing active-sterile vacuum oscillation solution to the
solar problem\cite{mm} 
can also lead to the equilibration of $\nu'_e$ if
$\delta m^2 > 10^{-7}$ eV$^2$.

Sterile neutrino models therefore have a cosmological credibility
problem. However, recent work has shown how a relic
neutrino-antineutrino asymmetry can be used to suppress active-sterile
transitions in the early universe\cite{fv}$^,$\cite{ftv}. 
This promises to meet the
cosmological challenge of BBN in full. It is quite possible, though
not definitively shown at the time of the present conference, that
large-angle active-sterile solutions to the solar and atmospheric
problems can be consistent with BBN through the agency of a
sufficiently large neutrino asymmetry $(\sim 10^{-5})$. I will review
this work in this talk. I first consider a pre-existing neutrino
asymmetry and determine the condition under which it can survive to
suppress active-sterile transitions prior to BBN.\cite{fv} 
I then consider the
much more interesting scenario whereby active-sterile neutrino
oscillations can themselves {\it create} the neutrino asymmetry that
then suppresses further such oscillations.\cite{ftv}

\section{Survival of pre-existing neutrino asymmetry}

Suppose that some unspecified mechanism creates a neutrino asymmetry $L$ 
at a temperature much higher than the epoch preceeding BBN ($> 100$ MeV,
say). The matter induced potential for an active species
$\nu_{\alpha}$ where $\alpha = e, \mu, \tau$ is then
\begin{equation}
V_{\alpha} = \sqrt{2} G_F n_{\gamma} \left( L^{(\alpha)} -
\frac{A_{\alpha}T^2}{m_W^2}\right),
\end{equation}
where $G_F$ is the Fermi constant, $A_e \simeq 55$,
$A_{\mu,\tau} \simeq 15.3$, $n_{\gamma}$ is the photon number
density, $m_W$ is the $W$ mass and
\begin{equation}
L^{\alpha} \simeq L_{\nu_{\alpha}} + L_{\nu_e} + L_{\nu_{\mu}} +
L_{\nu_{\tau}},
\end{equation}
with $L_f \equiv (n_f - n_{\bar{f}})/n_{\gamma}$ for fermion species
$f$. The matter mixing angle between $\nu_{\alpha}$ and a sterile
species $\nu'_{\alpha}$ is then
\begin{equation}
\sin^2 2 \theta_m = \frac{\sin^2 2 \theta_0}{1 - 2 z \cos 2 \theta_0 +
z^2}
\end{equation}
where $\theta_0$ is the vacuum mixing angle (which we for definiteness
set to be the maximal $\pi/4$ from now on) and
\begin{equation}
z \equiv - a + b \equiv \frac{2E}{\delta m^2} V_{\alpha},
\end{equation}
with $a$ being the neutrino asymmetry term and $b$ being the
$A_{\alpha}T^2/m_W^2$ term. At high $T$, the $b$ term dominates and
makes $\theta_m$ close to zero thus suppressing active-sterile
transitions. However, for interesting values of $\delta m^2$ the $b$
term ceases to be large enough at temperatures in the $5-30$ MeV
range. In this case, the asymmetry term $a$ needs to be sufficiently
large in order for the transitions to continue to be suppressed. We
need, roughly, that\cite{fv}
\begin{equation}
L^{(e)} > \frac{|\delta m^2|/eV^2}{400},\quad L^{(\mu,\tau)} >
\frac{|\delta m^2|/eV^2}{1600}.
\end{equation}

However, for the large mixing angles of interest it is well known that
the effect of the active-sterile oscillations is to destroy any
pre-existing $L^{(\alpha)}$. If the asymmetry is destroyed prior to
the $b$ term becoming too small, then subsequent oscillations will
bring $\nu'_{\alpha}$ into thermal equilibrium. The critical time is
during the MSW resonance when the matter mixing angle become maximal:
$\theta_m = \pi/4$. The condition that the asymmetry survive the
destructive tendencies of the MSW resonance is that\cite{fv}
\begin{equation}
L^{(\alpha)} > L^{(\alpha)}_{crit} \simeq \left(\frac{(\delta m^2)^4
m^2_P A^9_{\alpha}}{3\times 10^5 y^2_{\alpha} G^4_F m^{18}_W}
\right)^{1/11},
\label{crit}
\end{equation}
where $m_P$ is the Planck mass and $y_{\alpha}$ is a calculable
parameter that determines the collision frequency of $\nu_{\alpha}$
($y_e \simeq 4$, $y_{\mu,\tau} \simeq 2.9$). This condition arises by
demanding that the system pass through the MSW resonance briefly,
rather than being forced to track the MSW resonance as the resonance
temperature evolves to lower and lower temperatures as $L^{(\alpha)}$
is destroyed. So, a pre-existing asymmetry can suppress active-sterile
transitions provided it is sufficiently large. Evaluating 
Eq.(\ref{crit}) we get that\cite{fv}
\begin{equation}
L^{(e)} > 9 \times 10^{-6}\ for\ |\delta m^2| < 10^{-4} eV^2,\ \ 
L^{(\mu,\tau)} > 2 \times 10^{-5}\ for\ |\delta m^2| < 10^{-2} eV^2.
\end{equation} 

\section{Creation of neutrino asymmetry through oscillations} 

While the above is a good start, it is clear that a concrete mechanism
is needed for the creation of neutrino asymmetries. Remarkably,
active-sterile oscillations can themselves create such 
asymmetries!\cite{ftv}
Suppose that at some initial time or temperature the number
density of a sterile species is negligible compared to that of an
active species. One can show then that
\begin{equation}
\frac{d L_{\nu_{\alpha}}}{dt} \simeq \frac{3}{8}[ - \Gamma(\nu_{\alpha}
\to \nu'_{\alpha}) + \Gamma(\bar{\nu}_{\alpha} \to
\bar{\nu}'_{\alpha})],
\label{dLdt}
\end{equation}
where the transition rates are given by
\begin{equation}
\Gamma(\nu_{\alpha} \to \nu'_{\alpha}) \simeq \frac{1}{2} \sin^2 2\theta_m
\Gamma_{\alpha}
\end{equation}
and a similar expression for antineutrinos with the sign of the
asymmetry reversed.
The collision rate $\Gamma_{\alpha} = y_{\alpha} G_F^2 T^5$. A number
of other important assumptions are incorporated into the above: we
have assumed that the matter eigenstate basis diagonalises the
Hamiltonian, that the thermal spread of neutrino momentum can be
ignored, and that the mean free path for $\nu_{\alpha}$ is much larger
than the matter oscillation length. There is unfortunately not enough
room here to further discuss these approximations. Equation \ref{dLdt}
can then be reduced to\cite{ftv}
\begin{equation}
\frac{dL_{\nu_{\alpha}}}{dt} \simeq \frac{3 \Gamma_{\alpha} s^2 a (c - b)}
{4[1 - 2c(-a + b) + (a-b)^2][1 - 2c(a+b) + (a+b)^2]},
\label{dLdt2}
\end{equation}
where $s \equiv \sin\theta_0$ and $c \equiv \cos\theta_0$. We now do
{\it not} set $\theta_0$ to be $\pi/4$.

Consider the case where $\delta m^2 < 0$, so that $b > 0$. For high
$T$, $b > c$ and $dL/dt$ has the opposite sign to $L^{(\alpha)}$. This
means that the asymmetry will tend to be destroyed, a fact we used in
the previous section. However, there will a critical temperature when
$b = c$ given by
\begin{equation}
T_{crit} \simeq 13(16)\left(\frac{c|\delta m^2|}{eV^2}\right)^{1/6}
\end{equation}
for $\nu_e(\nu_{\mu,\tau})$. For temperatures below $T_{crit}$ the
magnitude of the asymmetry will be increased! Note that $L^{(\alpha)}
= 0$ is a fixed point that goes from being stable to unstable as the
crticial temperature is passed. Suppose that initially the asymmetry
is extremely small (but nonzero). Then at the critical temperature $a
\simeq 0$ and $b = c$, so that both neutrinos and antineutrinos hit
resonance simultaneously. Equation \ref{dLdt2} then has as an 
initial form $dL/dt$ proportional to $L$ 
and a brief period of rapid exponential growth ensues,
amplified by the simultaneous resonance conditions. The
asymmetry increases until $|a| \simeq c$ when the non-linearity in the
equation sets in. The system then settles down to approximately
track the $|a| \simeq c$ trajectory. This behaviour is illustrated in
an illuminating Figure that can be found in Ref.\cite{ftv}

There are now two important questions to ask: Can this effect be used
to construct a BBN-consistent large-angle active-sterile neutrino
model? Are the approximations used above valid? These are both very
profound questions, and the required analysis is complicated. A
preliminary calculation, presented at the conference, showed the
following: Consider a subsystem composed of $\nu_e$, $\nu_{\mu}$ and
$\nu'_e$. Small-angle mixing of $\nu_{\mu}$ with $\nu'_e$ can induce a
large muon-number asymmetry via the above mechanism. An analogous
calculation to that summarised in the previous section can then be
performed to show that large-angle  $\nu_e - \nu'_e$ 
oscillations will not destroy
this asymmetry provided that $|\delta m^2_{ee'}/eV^2| < 10^{-6}
|\delta m^2_{\mu e'}eV^2|^{11/12}$ 
adopting a self-evident notation of the
mass differences. The author and R. Foot have analysed the two
questions posed above in some depth since the conference. Our results
appear in Ref.\cite{forthcoming}.

\section*{Acknowledgments}
This work was supported by the Australian Research Council. The author
would like to thank Professor Matts Roos for organising a very
interesting conference. He is also grateful to Kari Enqvist and
Kimmo Kainulainen for some discussions that took place during the
conference.

\end{document}